\documentclass[aps,pra,twocolumn,amsmath,showpacs,letterpaper,floatfix]{revtex4}

\usepackage{amssymb}
\usepackage{graphicx}
\usepackage{grffile}
\usepackage{color}
\newcommand{\ket}[1]{|#1\rangle}
\newcommand{\bra}[1]{\langle#1|}

\begin{document}

\title{Statistical analysis of sampling methods in quantum tomography}
\author{Thomas Kiesel}
\affiliation{Arbeitsgruppe Quantenoptik, Institut f\"ur Physik, Universit\"at  Rostock, D-18051 Rostock,
Germany}
\begin{abstract}
	In quantum physics, all measured observables are subject to statistical uncertainties, which arise from the quantum nature as well as the experimental technique. We consider the statistical uncertainty of the so-called sampling method, in which one estimates the expectation value of a given observable by empirical means of suitable pattern functions. We show that if the observable can be written as a function of a single directly measurable operator, the variance of the estimate from the sampling method equals to the quantum mechanical one. In this sense, we say that the estimate is on the quantum mechanical level of uncertainty. In contrast, if the observable depends on non-commuting operators, e.g.~different quadratures, the quantum mechanical level of uncertainty is not achieved. The impact of the results on quantum tomography is discussed, and different approaches to quantum tomographic measurements are compared. It is shown explicitly for the estimation of quasiprobabilities of a quantum state, that balanced homodyne tomography does not operate on the quantum mechanical level of uncertainty, while the unbalanced homodyne detection does. 
\end{abstract}

\maketitle

\section{Introduction}

The theory of measurements in quantum physics has a long tradition. First, it requires a good understanding of the measurement devices. In quantum optics, the foundations for the photodetector have been established in the pioneering works of Mandel~\cite{PhotodetectorMandel}, Kelley and  Kleiner~\cite{PhotodetectorKelley} as well as Glauber~\cite{PhotodetectorGlauber}. On this basis, more elaborate measurement techniques have been developed, such as balanced homodyne detection~\cite{HomodyneDetection,HomodyneDetectionVogel}, unbalanced homodyne detection~\cite{Wallentowitz96} and eight-port homodyne detection schemes~\cite{EightPortScheme1,EightPortScheme2}. All mentioned schemes provide the possibility to collect information to completely characterize an arbitrary quantum state of light. 

Once, the experimental techniques are well-known, one needs suitable tomographic methods to convert the raw experimental data into a convenient representation of the quantum state. In case of balanced homodyne measurements, in which one records phase dependent quadratures, there are several approaches which have been proposed. In~\cite{VogelRisken}, the so-called inverse Radon transform is applied to estimate the Wigner function of the quantum state.  This approach can also be generalized to obtain different quasiprobability distributions, such as the Glauber-Sudarshan $P$ function~\cite{Kiesel08}. In order to calculate density matrices, one may choose between Fourier techniques~\cite{KuehnVogel}, direct sampling schemes~\cite{DAriano,DArianoLeonhardPaul,Zucchetti} or more involved maximum likelihood methods~\cite{MaxLike1,MaxLike2}. 

However, there has only been little interest in the statistical properties of quantum state estimation. The theoretical foundations have already been considered in~\cite{Holevo}, but in particular for quantum state tomography, the statistical uncertainties of the estimates are not examined deeply.  First examinations have been performed in~\cite{Banaszek_Uncertainty}. The advantage of the sampling approach is that it directly gives an simple estimate of the statistical uncertainty of the estimated quantity~\cite{DAriano2}, which we will shortly review below. In case of the maximum likelihood approach, statistical uncertainties can be evaluated by the inverse of the so-called Fisher information matrix, which requires some numerical effort~\cite{Rehacek}. Alternative methods for the estimation of uncertainties can also be found in~\cite{Scheel, DiGuglielmo}.

In the present work, we compare the statistical uncertainty, which arises in the direct sampling approach, to the quantum mechanical variance, which provides a lower bound set by the quantum nature of light. In Sec. II, we consider functions of a single observable, which can be directly measured, and show that the variance of the corresponding pattern function equals to the quantum mechanical one. In Sec. III, we examine observables for which complete quantum state tomography is required. Explicitly, we show that sampling from balanced homodyne detection data does not operate on the quantum mechanical level of uncertainty, and find that the statistical independence of quadratures at different phases is the reason. We also discuss the estimation of phase-space distributions in Sec. IV, and show that the unbalanced homodyne detection scheme provides estimates with quantum mechanical uncertainties. Section V is dedicated to an example in quantum state tomography to illustrate the impact of the results.

\section{Functions of a single observable}
\subsection{General considerations}
As a first step, let us assume that we observe a single physical quantity $\hat{A}$ and estimate the expectation value of a function of this observable, $\hat{F}(\hat{A})$. As $\hat{F}(\hat{A})$ is a Hermitian operator, it can be written in its spectral decomposition
\begin{equation}
	\hat{F}(\hat{A}) = \int_{\mathcal A} F(A) \ket A\bra A dA,\label{eq:spectral:decomp:F:A}
\end{equation}
where $\ket A$ is an eigenvector of $\hat{A}$ with the eigenvalue $A$. The set $\mathcal A$ is the set of all eigenvalues. In case of discrete eigenvalues, the integration has to be replaced by the corresponding sum. The eigenvectors can be chosen orthogonal, 
\begin{equation}
	\langle A | A'\rangle = \delta(A-A'),
\end{equation}
where the right hand side has to be interpreted as the Kronecker symbol in the case of discrete eigenvalues. Furthermore, the eigenstates provide a resolution of identity,
\begin{equation}
	\int_{\mathcal A} \ket A\bra A dA = \hat{1}.
\end{equation}

Let us now consider an experiment which records a set of $N$ eigenvalues $\{A_j\}_{j=1}^N$ from $\mathcal A$ as outcomes. The underlying quantum state shall be denoted by $\hat{\rho}$. Then, the \emph{empirical expectation value} of some function $F(A)$ can be estimated as
\begin{equation}
	\tilde{F} = \frac{1}{N}\sum_{j=1}^N F(A_j).\label{eq:sampling:F:A}
\end{equation}
The tilde indicates that this quantity is a random number, since it is obtained from measured (and therefore random) values $\{A_j\}_{j=1}^N$, whose probabilities shall be denoted by $p(A_j)$. Therefore, the expectation value of this random number $F$ is given by
\begin{equation}
	\overline{\tilde{F}} = \int_{\mathcal A} p(A) F(A) dA.\label{eq:expect:sampling:F:A}
\end{equation}
As above, the integral has to be seen as a sum over the probabilities if the set of eigenvalues is discrete.
The key point is now that Eq.~\eqref{eq:sampling:F:A} provides a good estimate for the \emph{quantum mechanical expectation value} of the operator $\hat{F}(\hat{A})$. The probabilities of the outcomes are connected to the underlying quantum state by
\begin{equation}
	p(A) = {\rm Tr}\{\hat{\rho}\ket A\bra A\}.
\end{equation}
Inserting this into Eq.~\eqref{eq:expect:sampling:F:A} and applying~\eqref{eq:spectral:decomp:F:A}, we find 
\begin{equation}
	\overline{\tilde{F}} = {\rm Tr}\{\hat{\rho} \hat{F}(\hat{A})\}.\label{eq:equivalence:of:expectations}
\end{equation}
The Eqs.~\eqref{eq:sampling:F:A},\eqref{eq:expect:sampling:F:A} and~\eqref{eq:equivalence:of:expectations} form the basis of the sampling technique. In order to find an unbiased estimate of the expectation value of $\hat{F}(\hat{A})$, one simply has to insert his measurement outcomes $\{A_j\}_{j=1}^N$ into the so-called pattern function $F(A)$ and calculate the empirical mean $\tilde{F}$ of the values according to Eq.~\eqref{eq:sampling:F:A}. 

For the ability to make justified statements, one still needs a measure of the uncertainty of the estimate~\eqref{eq:sampling:F:A}. The \emph{empirical variance of the sampling points} $\{F(A_j)\}_{j=1}^N$ is given by
\begin{equation}
	\sigma^2_{F(A)} = \frac{1}{N-1}\sum_{j=1}^N (F(A_j) - \tilde{F})^2.\label{eq:empirical:variance:F:A}
\end{equation}
This number quantifies the spreading of the points $F(A_j)$. The factor $\frac{1}{N-1}$ guarantees that the estimate is unbiased, i.e.
\begin{equation}
	\overline{\sigma^2_{F(A)}} = \int_{\mathcal A} p(A) \left[F(A) - \overline{\tilde{F}}\right]^2 dA.
\end{equation}
Practically, the estimation of the empirical variance requires the calculation of the second moment of $F(A)$,
\begin{equation}
	\tilde{F}^{(2)} = \frac{1}{N}\sum_{j=1}^N F(A_j)^2.
\end{equation}
We easily see that this is exactly the sampling equation~\eqref{eq:sampling:F:A}, just with the square of the function $F(A)$. Therefore, and due to the orthogonality of the eigenstates $\ket A$, its expectation value again equals to the quantum mechanical one,
\begin{equation}
	\overline{\tilde{F}^{(2)}} = {\rm Tr}\{\hat{\rho} \hat{F}^2(\hat{A})\}.
\end{equation}
The same holds for arbitrary moments of the function $F(A)$.
As a consequence, also the \emph{quantum mechanical variance} of $\hat{F}(\hat{A})$, 
\begin{equation}
	\langle(\Delta \hat{F}(\hat{A}))^2\rangle = {\rm Tr}\{\hat{\rho}[\hat{F}(\hat{A})]^2\} - [{\rm Tr}\{\hat{\rho}\hat{F}(\hat{A})\}]^2
\end{equation}
can be estimated without bias from the set of sampling points $\{F(A_j)\}_{j=1}^N$:
\begin{equation}
	\overline{\sigma^2_{F(A)}} = \langle(\Delta \hat{F}(\hat{A}))^2\rangle
\end{equation}
Finally, we are interested in the statistical uncertainty on the estimate~\eqref{eq:sampling:F:A}. From estimation theory, this is just the empirical variance $\sigma^2_{F(A)}$ of the sampling points , divided by the number of measurements:
\begin{equation}
	\sigma^2_{\tilde{F}} = \sigma^2_{F(A)} / N.
\end{equation}
The factor $1/N$ guarantees that for an increasing number of data points $N$, the uncertainty of the empirical mean value is decreasing. For $N\to\infty$, the latter approaches stochastically the quantum mechanical expectation value.

In conclusion, we may estimate the quantum mechanical expectation value of the operator $\hat{F}(\hat{A})$ by the sampling equation~\eqref{eq:sampling:F:A}, and the variance of $\tilde{F}$ is exactly  expected to match the quantum mechanical variance $\langle(\Delta \hat{F}(\hat{A}))^2\rangle$, divided by the number of data points. In this sense, we may state that the determination of ${\rm Tr}\{\hat{\rho} \hat{F}(\hat{A})\}$ can be done on the quantum mechanical level of uncertainty. There are no other sources of noise contributing to the uncertainty, and it is not possible to achieve less fluctuations with classical statistical means.

 We emphasize that so far we only considered a function of the directly measurable operator $\hat{A}$, making this result possible. After briefly discussing some examples, we show that the situation becomes completely different when we require several non-commuting observables to estimate the expectation value of an operator $\hat{F}$.

\subsection{Functions of a single quadrature}

As a first example, let us consider the measurement of the quadrature operator $\hat x$, which is frequently done in balanced homodyne measurements. In this case, the set of eigenvalues is given by the continuous spectrum $\mathcal A = \mathbb R$. According to the above calculations, any function of a single quadrature can be estimated at the quantum mechanical level of uncertainty. This includes all kinds of moments of the quadrature, for instance normally ordered ones. However, note that we only may use a quadrature at a single phase. If we consider functions of quadratures of different phases, the situation will become completely different, as we will show below.
 
\subsection{Functions of the photon number}

Photon number resolving detectors can record the outcomes of the photon number operator $\hat n$. Here, the set of eigenvalues is the discrete spectrum $\mathcal A = \mathbb N$. Again, we can state that the expectation value of any function of the photon number operator can be estimated at the quantum mechanical level of uncertainty. We still note that in practice one can only record a finite number of measurements, leading to some uncertainty on photon number probabilities with very low values, which typically occur for very large photon numbers. However, this problem can be minimized by increasing the number of measurements.

\section{Expectation values from balanced homodyne detection}

Measurements of a single operator, such as the quadrature or the photon number, can not characterize a quantum state completely. For some operators, one needs more information about the state in order to estimate some expectation value. For instance, measurements of the quadrature distributions for all phases in $[0,\pi)$ are informationally complete~\cite{Prugovecki,Busch}, and we may calculate any expectation value from this measurement outcomes. On the other hand, it has been shown that so-called quasiprobability representations of quantum states can be retrieved by photon number resolved measurements, when one displaces the quantum state in phase space before the measurement~\cite{Wallentowitz96}. In the following, we consider these methods more in detail.

\subsection{Quantum mechanical expectation values}

For having a meaningful reference, we start with the calculation of the quantum mechanical variance of some operator $\hat{F}$. Here, we express it in terms of the characteristic functions of the Wigner function of the density operator $\hat{\rho}$ of the state and of the observable $\hat{F}$. In general, the characteristic function of the Wigner function is defined as
\begin{equation}
	\Phi_{\hat{F}}(\beta) = {\rm Tr}\{\hat{F} \hat D(\beta)\}, \label{eq:define:CF}
\end{equation}
with $\hat D(\beta) = e^{\beta\hat a^\dagger-\beta^*\hat a}$ being the well-known displacement operator~\cite{Perelomov}. For Hermitian operators, we have the relation $\Phi_{\hat{F}}(-\beta) = \Phi^*_{\hat{F}}(\beta)$. Moreover, if $\hat{F}$ is the density operator of a state, we will omit the index of $\Phi(\beta)$ throughout the paper. Conversely, the operator may be retrieved from its characteristic function by
\begin{equation}
	\hat{F} = \frac{1}{\pi} \int d^2\beta\, \Phi^*_{\hat{F}}(\beta) \hat D(\beta).\label{eq:Phi:to:rho}
\end{equation}
With these quantities at hand, we may calculate the expectation value of $\hat{F}$ with respect to the state $\hat{\rho}$ by
\begin{equation}
	{\rm Tr}\{\hat{\rho}\hat{F}\} = \frac{1}{\pi}\int d^2\beta\,\Phi(\beta) \Phi^*_{\hat{F}}(\beta).\label{eq:expect:with:Phi}
\end{equation}
Let us now express the second moment of $\hat{F}$ in terms of characteristic functions. Inserting Eq.~\eqref{eq:Phi:to:rho} for the operator $\hat{F}$, we obtain
\begin{eqnarray}
	{\rm Tr}\{\hat{\rho}\hat{F}^2\} &=& \frac{1}{\pi^2}\int d^2\beta'\int d^2\beta'' \Phi_{\hat{F}}^*(\beta')\Phi_{\hat{F}}^*(\beta'')	\nonumber\\&&\quad\times{\rm Tr}\{ \hat{\rho}\hat D(\beta')\hat D(\beta'')\}.
\end{eqnarray}
By applying the equality
\begin{equation}
	\hat D(\alpha)\hat D(\beta) = \hat D(\alpha+\beta) e^{i {\rm Im}(\alpha\beta^*)},
\end{equation}
and writing both integrals in polar coordinates, we find the final expression
\begin{eqnarray}
	{\rm Tr}\{\hat{\rho}\hat{F}^2\} &=& \frac{1}{\pi^2}\int d\varphi\, d\phi\, db'\, db''\,|b'||b''| \Phi(b'e^{i\varphi}+b''e^{i\phi})\nonumber\\&&\times\Phi^*_{\hat{F}}(b'e^{i\varphi})\Phi^*_{\hat{F}}(b''e^{i\phi}) e^{ib'b''\sin(\varphi-\phi)}.\label{eq:expect:hat:F2}
\end{eqnarray}
In our notation, the integrals over the angles $\varphi,\phi$ range from $0$ to $\pi$, while the integration over $b',b''$ covers the full real line. The variance can then be easily calculated as
\begin{equation}
	{\rm Tr}\{(\Delta\hat{F})^2\} = {\rm Tr}\{\hat{\rho}\hat{F}^2\} - [{\rm Tr}\{\hat{\rho}\hat{F}\}]^2.\label{eq:var:F}
\end{equation}
Equation~\eqref{eq:expect:hat:F2} will be the reference for comparison with the variance arising in sampling methods.

\subsection{Balanced homodyne measurements}

\subsubsection{Sampling from balanced homodyne measurement data}
Let us assume that the state $\hat{\rho}$ is sent to a balanced homodyne detector~\cite{HomodyneDetection,HomodyneDetectionVogel}, recording quadrature values $\{x_j,\varphi_j)\}_{j=1}^N$ at different phases $\varphi_j \in [0,\pi)$. Here, we assume that the phase values are uniformly distributed within this interval. Then, the quadrature $x_j$ follows the quadrature distribution $p(x_j;\varphi_j)$, which is conditioned on the value of $\varphi_j$. The joint probability distribution is given by $p(x;\varphi)/\pi$.

Sampling is an established technique to estimate the expectation value of an Hermitian operator $\hat{F}$ from this set of data by an empirical mean of a suitable pattern function $f(x, \varphi)$,
\begin{equation}
	\tilde{F} = \frac{1}{N}\sum_{j=1}^N f(x_j,\varphi_j).\label{eq:estimate:F:x:phi}
\end{equation}
Analogously to Eq.~\eqref{eq:sampling:F:A}, this number $\tilde{F}$ is a random variable, whose expectation value is given by
\begin{equation}
	\overline{\tilde{F}} = \int_{-\infty}^\infty dx \int_0^\pi d\varphi\, \frac{p(x; \varphi)}{\pi} f(x, \varphi).
	\label{eq:expect:f(x,phi)}
\end{equation}
The pattern function has to be designed such that this expectation value equals to the quantum mechanical one,
\begin{equation}
	\overline{\tilde{F}} = {\rm Tr}\{\hat{\rho}\hat{F}\}.
	\label{eq:expect:with:pattern}
\end{equation}
Let us now find such a suitable pattern function belonging to the operator $\hat{F}$. The characteristic function of the state is connected to the quadrature distribution as
\begin{equation}
 	\Phi(\beta) = \int_{-\infty}^\infty dx\, p\left(x;\arg(\beta)-\tfrac{\pi}{2}\right) e^{i|\beta| x}. \label{eq:p(x):to:Phi}
\end{equation}
Inserting this equation into Eq.~\eqref{eq:expect:with:Phi} and writing the integration over $\beta$ in polar coordinates, we obtain
\begin{eqnarray}
	{\rm Tr}\{\hat{\rho}\hat{F}\} &=& \int_{-\infty}^\infty dx \int_0^\pi d\varphi \frac{p\left(x;\varphi\right)}{\pi}\nonumber\\&&\times\int_{-\infty}^\infty db\, |b|\,  e^{i b x} \Phi^*_{\hat{F}}(i b e^{i\varphi}).
\end{eqnarray}
From this relation, we easily see that the pattern function is given by
\begin{equation}
	f(x,\varphi) = \int_{-\infty}^\infty db\, |b|\,  e^{ib x} \Phi^*_{\hat{F}}(i b e^{i\varphi}).\label{eq:define:pattern:function}
\end{equation}
By construction, the expectation value of this pattern function with respect to the joint quadrature distributions $p(x;\varphi)/\pi$ always equals to the quantum mechanical expectation value, see Eq.~\eqref{eq:expect:with:pattern}. 

Moreover, we are interested in the empirical variance of the single data points, which can be estimated from the experimental data by
\begin{equation}
	\sigma^2_{f(x,\varphi)} = \frac{1}{N-1}\sum_{j=1}^N(f(x_j,\varphi_j) - \tilde{F})^2,\label{eq:empirical:variance:f(x:phi)}
\end{equation}
being completely analogous to Eq.~\eqref{eq:empirical:variance:F:A}. Consequently, the expectation value of this variance is given by
\begin{equation}
	\overline{\sigma^2_{f(x,\varphi)}} = \overline{[f(x,\varphi)]^2} - \overline{f(x,\varphi)}^2.\label{eq:pattern:variance}
\end{equation}
Finally, the variance of the estimate $\tilde F$ for the expectation value of the operator $\hat F$ can be obtained from
\begin{equation}
	\sigma^2_{\tilde{F}} = \sigma^2_{f(x,\varphi)} / N.
\end{equation}
This number quantifies the uncertainty of the estimate~\eqref{eq:estimate:F:x:phi}.

Let us now examine if we can expect the empirical variance Eq.~\eqref{eq:pattern:variance}  to match the quantum mechanical variance of the operator $\hat F$. Contrarily to the procedure above, the operator $\hat F$ now depends on quadratures at different phases $\varphi$, whose operators do not commute. For getting a deeper understanding, we concentrate on the second moment of $f(x,\varphi)$. By using the inverse relation of Eq.~\eqref{eq:p(x):to:Phi}, 
\begin{equation}
	p(x;\varphi) = \frac{1}{2\pi} \int_{-\infty}^\infty db\, e^{-ibx} \Phi(i b e^{i\varphi}),
\end{equation}
we find 
\begin{eqnarray}
	\overline{f(x,\varphi)^2} &=& \int_{-\infty}^\infty dx \int_0^\pi d\varphi\, \frac{p(x; \varphi)}{\pi} [f(x,\varphi)]^2\\
	&=& \frac{1}{2\pi^2} \int dx\, d\varphi\, db\, db'\, db''\,|b'||b''| e^{i(b'+b''-b)x}\nonumber\\&&\times \Phi(i b e^{i\varphi})\Phi^*_{\hat{F}}(i b' e^{i\varphi})
	 \Phi^*_{\hat{F}}(i b'' e^{i\varphi}).\label{eq:expect:f2:1}
\end{eqnarray}
We substitute $\varphi\to\varphi - \pi/2$ in order to remove the imaginary unit $i$ in the arguments of the characteristic function. A careful analysis shows that we do not have to change the integration domain due to the periodicity of the integrand. Moreover, the integral over $x$ can be evaluated as
\begin{equation}
	\frac{1}{2\pi} \int_{-\infty}^\infty dx\, e^{i(b'+b''-b)x} = \delta(b'+b''-b),
\end{equation}
which can be used to evaluate another integral in \eqref{eq:expect:f2:1}:
\begin{eqnarray}
	\overline{f(x,\varphi)^2} &=& \frac{1}{\pi}\int d\varphi\, db'\, db''\,|b'||b''| \Phi((b' + b'') e^{i\varphi})\nonumber\\&&\times \Phi^*_{\hat{F}}(b' e^{i\varphi})
	 \Phi^*_{\hat{F}}(b'' e^{i\varphi}).\label{eq:expect:f2:2}
\end{eqnarray}
Together with Eq.~\eqref{eq:pattern:variance}, we find the theoretically expected variance of the sampling method.

We stress that this equation is not directly evaluated in practice, since the underlying quadrature distribution is unknown, and we only have a sample of quadrature measurements $\{(x_j,\varphi_j)\}_{j=1}^N$. Instead, we use the empirical variance given in Eq.~\eqref{eq:empirical:variance:f(x:phi)}. However, the theoretical expectation is given by Eq.~\eqref{eq:pattern:variance} in combination with Eq.~\eqref{eq:expect:f(x,phi)} and \eqref{eq:expect:f2:2} and the basis for all following considerations.

\subsubsection{Comparison with the quantum mechanical variance}

Let us compare the quantum mechanical variance~\eqref{eq:var:F} with the variance expected from the sampling method~\eqref{eq:empirical:variance:f(x:phi)}.  Since the first moments of the sampling method and quantum mechanics are equal by construction, it is sufficient to examine the second moments of $f(x;\varphi)$ and $\hat{F}$. The expressions from the sampling method \eqref{eq:expect:f2:2} and the quantum mechanical calculation \eqref{eq:expect:hat:F2} look quite similar, but they are different: in \eqref{eq:expect:f2:2}, one integration over $\phi$ is missing. A closer look reveals that if one replaces 
\begin{equation}
	\frac{1}{\pi}\int_0^\pi d\phi\rightarrow \int d\phi\delta(\varphi-\phi)
\end{equation}
in the quantum mechanical expectation \eqref{eq:expect:hat:F2}, one finds the expression for the pattern function \eqref{eq:expect:f2:2}. Obviously, since we only observe the quadrature distribution $p(x;\varphi)$ at a fixed phase $\varphi$, and all measured quadratures for different $\varphi$ are stochastically independent, the ``correlations'' of the quadrature distributions between different phases $\varphi$ and $\phi$ are not taken into consideration. This is due to the fact that $p(x;\varphi)$ is not the joint distribution of all quadratures, whose  marginals are the observed quadrature distributions. The definition of the joint distribution suffers from the problem of the non-commutativity of the corresponding quadrature operators and is closely related to the different phase-space distributions.  As a consequence, the quantum mechanical variance is not equal to the empirical variance of the pattern function. In this sense, the statistics of the balanced homodyne measurements is not at the quantum mechanical level, since the expectation value of $\hat{F}$ can not be estimated with quantum mechanical uncertainty. 

The definition of a joint probability of two quadratures
suffers from the problem of the non-commutativity of two quadratures for
different phases. In this context there appear similar problems as 
the well known ambiguity of the definition of phase-space distributions.

We also note that the statement does not change when the examined operator is phase-insensitive, i.e.~
\begin{equation}
	\Phi_{\hat{F}}(b e^{i\varphi}) \equiv \Phi_{\hat{F}}(b).
\end{equation}
In this case, it is sufficient to record the completely phase-diffused quadrature distribution 
\begin{equation}
	p_{\rm pd}(x) = \frac{1}{\pi}\int p(x; \varphi) d\varphi, \label{eq:phase:diffused:quad:dist}
\end{equation}
which can be easily done by choosing a uniform phase distribution of the local oscillator coherent state. This scheme may be practically more simple, since one does not have to record the phase values. However, it is still necessary to guarantee the uniform phase distribution in~\eqref{eq:phase:diffused:quad:dist}. Moreover, it does not bring any statistical benefit, since the variance~\eqref{eq:expect:f2:2} of the pattern function remains the same. Therefore, the same number of data points is required to obtain the same statistical precision.

Finally, we might ask if it is possible to experimentally construct a bipartite state described by the characteristic function $\Phi(\beta'+\beta'')$, which appears in Eq.~\eqref{eq:expect:hat:F2}. The arguments $\beta'$ and $\beta''$ are assigned to each of the two subsystems. If this was possible, one could use it for estimating the quantum mechanical variance from Eq.~\eqref{eq:expect:hat:F2}. Practically, we would need to measure joint quadrature distributions 
\begin{eqnarray}
	p(x_1,x_2;\varphi_1,\varphi_2) &=& \frac{1}{(2\pi)^2} \int_{-\infty}^\infty db'\int_{-\infty}^\infty db'' e^{-(i b'x_1+ b''x_2)} \nonumber\\&&\qquad\times\Phi(ib' e^{i\varphi_1} + ib'' e^{i\varphi_2}),
\end{eqnarray}
which seems to require joint balanced homodyne measurements on the bipartite state.

The problem is just that $\Phi(\beta'+\beta'')$ does not refer to a physical state. This becomes clear when we examine the covariance matrix. The required moments can be obtained by taking the derivatives of $\Phi(\beta'+\beta'')$ at $\beta'=\beta''=0$. We note that all moments of the bipartite state can be expressed in moments of the state $\Phi(\beta)$, since 
\begin{eqnarray}
	&&\frac{\partial^k}{\partial \beta'^k}\frac{\partial^l}{\partial (\beta'^*)^l}\frac{\partial^m}{\partial \beta''^m}\frac{\partial^n}{\partial (\beta''^*)^n}\Phi(\beta'+\beta'')\big|_{\beta',\beta''=0} \nonumber\\
		&&\qquad = \frac{\partial^{k+m}}{\partial \beta^{k+m}}\frac{\partial^{l+n}}{\partial (\beta^*)^{l+n}}\Phi(\beta)\big|_{\beta=0}.
\end{eqnarray}
Therefore, if the state described by $\Phi(\beta)$ has a quadrature covariance matrix 
\begin{equation}
	C_1 = \left(\begin{array}{c c} V_x & C_{xp}\\ C_{xp} & V_p\end{array}\right),
\end{equation}
the bipartite state can be characterized by the matrix 
\begin{equation}
	C_2 = \left(\begin{array}{c c } C_1 & C_1\\ C_1 & C_1\end{array}\right).
\end{equation}
In order to describe a physical state, this matrix has to satisfy the nonnegativity condition~\cite{CovarianceOfState}
\begin{equation}
	C_2  + i \Omega \geq 0,\label{eq:cond:on:cov}
\end{equation}
where 
\begin{equation}
	\Omega = \left(\begin{array}{c c} J & 0\\ 0 & J\end{array}\right) \quad \mbox{and} \quad 
		J = \left(\begin{array}{c c} 0 & 1\\ -1 & 0\end{array}\right).
\end{equation}
However, the $3\times3$ minors of $C_2$ are always negative, e.g.
\begin{equation}
	\left|\begin{array}{c c c} V_x 		& C_{xp}+i & V_x\\
							C_{xp}-i 	& V_p		& C_{xp}\\
							V_x			& C_{xp}	& V_x\end{array}\right| = - V_x.
\end{equation}
Therefore, Eq.~\eqref{eq:cond:on:cov} is violated by the covariance matrix $C_2$ of the bipartite characteristic function $\Phi(\beta'+\beta'')$, and $\Phi(\beta'+\beta'')$ can never correspond to a physical quantum state, which could be generated in an experiment. As a consequence, it seems unfeasible to estimate the variance~\eqref{eq:expect:hat:F2} on the quantum mechanical level. 


\subsubsection{Coherent displacement of states and operators.}\label{sec:displacement:and:operators}

Next, we consider a more general setting. First, we assume that we look at a family of observables $F(\gamma)$, being constructed by a coherent displacement of some initial operator $\hat{F}$:
\begin{equation}
	\hat{F}(\gamma) = \hat D(\gamma) \hat{F} \hat D(-\gamma).
\end{equation}
The characteristic function of these operators is given by
\begin{equation}
	\Phi_{\hat{F}(\gamma)} (\beta) = \Phi_{\hat{F}}(\beta) e^{\beta\gamma^*-\beta^*\gamma}.\label{eq:Phi:F(alpha)}
\end{equation}
\begin{figure}
	\includegraphics[width=0.6\columnwidth]{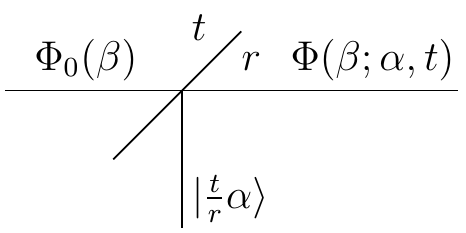}
	\caption{Scheme for overlapping a signal state $\Phi_0(\beta)$ with a coherent state.}
	\label{fig:setup}
\end{figure}
Furthermore, we also apply the displacement to the state $\hat{\rho}$, but we look at a suitable experimental realization. Let us send the initial state, described by $\Phi_0(\beta)$, to a beamsplitter with real transmissivity $t$ and reflectivity $r$, satisfying $t^2+r^2=1$ (see Fig.~\ref{fig:setup}). At the second input of the beamsplitter, we irradiate a coherent state with amplitude $t\alpha/r$, described by its characteristic function
\begin{equation}
	\Phi_{\rm coh} (\beta) = e^{(t \alpha^*/r)\beta -(t \alpha/r)\beta^*} e^{-|\beta|^2/2}.
\end{equation}
 The resulting output state has the characteristic function
\begin{eqnarray}
	\Phi(\beta; \alpha, t) &=&
		\Phi_0(t\beta) \Phi_{\rm coh}(r\beta)\\
		&=&\Phi_0(t\beta) e^{t\alpha^*\beta-t\alpha\beta^*} e^{-(1-t^2)|\beta|^2/2}.\label{eq:mix:input:with:coh}
\end{eqnarray}
We have chosen the amplitude of the coherent state such that the reflectivity does not appear in all calculations. Note that the transmissivity $t$ can also be used for taking the detector quantum efficiency into account. It is well known that an imperfect detector with efficiency $\eta < 1$ can be modelled by first mixing the input state with a fraction of $1-\eta$ of vacuum and subsequently performing the measurement with an ideal detector. The transmissivity of the corresponding beamsplitter is given by $\sqrt{\eta}$. By applying Eq.~(\ref{eq:mix:input:with:coh}) on the state $\Phi(\beta; \alpha, t)$ with transmissivity $t' = \sqrt{\eta}$ and coherent state amplitude $\alpha' = 0$, we find
\begin{eqnarray}
	\Phi_\eta(\beta;\alpha,t) &=& \Phi(\sqrt{\eta}\beta)  e^{-(1-\eta)|\beta|^2/2}\nonumber\\
			&=&\Phi(\beta; \alpha, t\sqrt{\eta}).
\end{eqnarray}
Therefore, an imperfect detector can be simply taken into account by replacing the transmissivity $t$ of the beam splitter by the effective transmissivity $t\sqrt{\eta}$. In consequence, we may consider only ideal detectors.

Of course, the expectation value of the pattern function corresponding to $\hat{F}(\alpha)$ with respect to the new output state is exactly the quantum mechanical expectation again. Let us look at the second moment of the pattern function, by inserting Eq.~\eqref{eq:Phi:F(alpha)} and \eqref{eq:mix:input:with:coh} into \eqref{eq:expect:f2:2}:
\begin{widetext}
	\begin{eqnarray}
		\overline{f(x,\varphi)^2} &=& \frac{1}{\pi}\int_0^\pi d\varphi\int_{-\infty}^\infty db'\, \int_{-\infty}^\infty db''\,|b'||b''|\, \Phi_0(t(b' + b'')e^{i\varphi})e^{-(1-t^2)(b'+b'')^2/2}
		\nonumber\\ &&\qquad\qquad \times\Phi^*_{\hat{F}}(b' e^{i\varphi})
	 \Phi^*_{\hat{F}}(b'' e^{i\varphi})
		e^{(b'+b'')e^{i\varphi}(t\alpha^* -\gamma^*) - (b'+b'')e^{-i\varphi}(t \alpha-\gamma)}.\label{eq:expect:f2:full}
	\end{eqnarray}
\end{widetext}
We focus on the following aspect: We want to estimate the quantum mechanical expectation value of the operator $\hat{F}(\gamma)$ with respect to the initial state. For this purpose, we have two possibilities:
\begin{enumerate}
 \item We do not displace the initial state and omit the beamsplitter. Mathematically, this is given by $t=1$ and $\alpha = 0$. The expectation value of $\hat{F}(\gamma)$ is obtained by choosing the pattern function corresponding to $\hat{F}(\gamma)$, i.e.~by suitable calculations after the balanced homodyne measurement.
	\item We displace the state by $\alpha = -\gamma/t$. The calculations after the measurement only require the pattern function for $\hat{F}$, i.e.~we set $\gamma = 0$ in Eq.~\eqref{eq:expect:f2:full}. 
\end{enumerate}
The expectation value of the sampling procedure  is in both cases the same, namely the quantum mechanical expectation. However, the uncertainties differ: In the second case, the initial state enters as $\Phi_0(t\beta)$, which is the state after exposition to losses with $\eta = t^2$. Therefore, we may expect a worse result than in the first scheme, which only depends on the perfect state. This finding also holds when we consider imperfect detection: If the quantum efficiency of the balanced homodyne detector equals to $\eta_d$, we had to replace $t\to\sqrt{\eta_d}$ in the first case and $t\to t\sqrt{\eta_d}$ in the second case. Therefore, the beamsplitter which displaces the initial state introduces unavoidable losses. This has consequences for the discussion of quantum state tomography.

\section{Relation to quantum state tomography}

In the case of quantum state tomography, one is interested to find a complete representation of the quantum state. Here, we discuss the representation with means of quasiprobabilities. In many cases, these quasiprobabilities can be represented as the expectation value of a displaced operator $\hat{F}$, which is phase-insensitive:
\begin{equation}
	P(\alpha) = {\rm Tr}\{\hat{\rho} \hat D(\alpha) \hat{F}\hat D(-\alpha)\}.\label{eq:P(alpha)}
\end{equation}
The coefficients $F_n$ in the Fock basis expansion of $\hat{F}$,
\begin{equation}
	\hat{F} = \sum_{n=0}^\infty F_n \ket n\bra n,
\end{equation}
determine a specific quasiprobability. For instance, the Wigner function is obtained by choosing $F_n = 2/\pi (-1)^n$, while the $Q$ function arises from $F_n = \delta_{n,0}/\pi$. More generally, the coefficients of $\Omega$-ordered quasiprobabilities~\cite{AgarwalWolf} can be written as
\begin{equation}
	\Omega_n = \frac{2}{\pi}\int_0^\infty d b\, b\, \Omega_{w}(b)  L_n(b^2),
\end{equation}
with $L_n(x)$ being the $n$th Laguerre-polynomial. For the family of $s$-parameterized quasiprobabilities, one chooses $\Omega(b;s) = e^{-(1-s) b^2/2}$, for nonclassicality quasiprobabilities one uses a suitable nonclassicality filter~\cite{Kiesel10,Witnesses}.

We have discussed different techniques for estimating expectation values of the form~\eqref{eq:P(alpha)}. 
First, we notice that the so-called cascaded balanced homodyning technique~\cite{Kis99} does not provide any advantage over the standard balanced homodyne detection. The former corresponds to the method 2 described in the previous section, where one displaces the state and estimates $\hat{F}$ for $\gamma = 0$, while the latter is realized by the method 1. Obviously, the former suffers a reduction of the quantum efficiency by the transmissivity of the first beamsplitter. This is not affected by the fact that $\hat{F}$ is phase-independent, and recording of the phase values is not required in the cascaded measurement. One can only improve the situation by choosing a beamsplitter with high transmissivity.

Second, we note that balanced homodyne measurements followed by sampling of quasiprobabilities does not work on the quantum mechanical level of uncertainty. Therefore, we can not expect that this scheme is optimal for this purpose. Indeed, if one is interested in quasiprobabilities, there is a better alternative: The unbalanced homodyne detection technique is based on a different interpretation of Eq.~\eqref{eq:P(alpha)}: First, the quantum state is displaced in the same way described in Sec.~\ref{sec:displacement:and:operators}. Afterwards, the expectation value of the phase-independent operator $\hat{F}$ is sampled from photon number measurements. Since this only requires the measurement of a single observable, namely the photon number, the estimation of the quasiprobability at a specific point $\alpha$ is performed on the quantum mechanical level of uncertainty. Provided that the balanced homodyne detector and the photon number resolved  detector had the same quantum efficiency, we recommend to choose the latter one, since we expect it to give better results.

Finally, we emphasize that the unbalanced scheme is optimal for the estimation of the phase-space representation at fixed points $\alpha$, but does not cover correlations between different points $\alpha,\alpha'$. Therefore, if one wants to estimate quantities which require the knowledge of the quasiprobability at different $\alpha$, one can not expect to achieve this at the quantum mechanical level of uncertainty as well. For instance, quadrature distributions are better measured in balanced homodyne detection schemes. In this sense, the unbalanced measurement is optimal for the estimation of quasiprobabilities, but not always the best in different cases.

\section{Example}

To demonstrate the difference of the statistical uncertainty in balanced homodyne and photon-number resolved detection, let us consider the determination of a nonclassicality quasiprobability of a squeezed state~\cite{Kiesel10,Kiesel11}. Its variances are $V_x = 0.5$ and $V_p = 2.0$, whereas the variance of the vacuum state shall be $V_{\rm vac} = 1$. We will apply 
\begin{equation}
	\Omega_w(\beta) = \int \omega(\beta')\omega(\beta'+\beta/w) d^2\beta
\end{equation}
as a filter, with $\omega(\beta) = (2/\pi)^{3/4} e^{-|\beta|^4}$. The prefactor guarantees that $\Omega_w(0) = 1$, and the filter width is fixed with $w = 1.8$. Moreover, the pattern function~\eqref{eq:define:pattern:function} is defined by choosing $\Phi_{\hat{F}}(\beta) = \pi^{-1}\Omega_w(\beta) e^{|\beta|^2/2}e^{\alpha^*\beta-\alpha\beta^*}$.

For the balanced homodyne detection scheme, we generate a set of $N = 100000$ data points, each consisting of a pair $(\varphi_j, x_j)$. The phase values $\varphi_j$ are uniformly distributed in the interval $[0,\pi)$, whereas  $x_j$ follows the quadrature distribution $p(x;\varphi_j)$ conditioned on the value of $\varphi_j$. From this simulated set of data, we sample the nonclassicality quasiprobability together with its statistical uncertainty. For reasons of simplicity, we assume to have an ideal detector, i.e.~$\eta = 1$. 

In case of the unbalanced homodyne detection scheme, we calculate the photon-number distribution $p_n$ together with its variance theoretically, the maximum photon number is restricted to $n = 20$. Then we derive the statistical uncertainty from the result by means of linear error propagation. 

\begin{figure}
	\includegraphics[width=0.8\columnwidth]{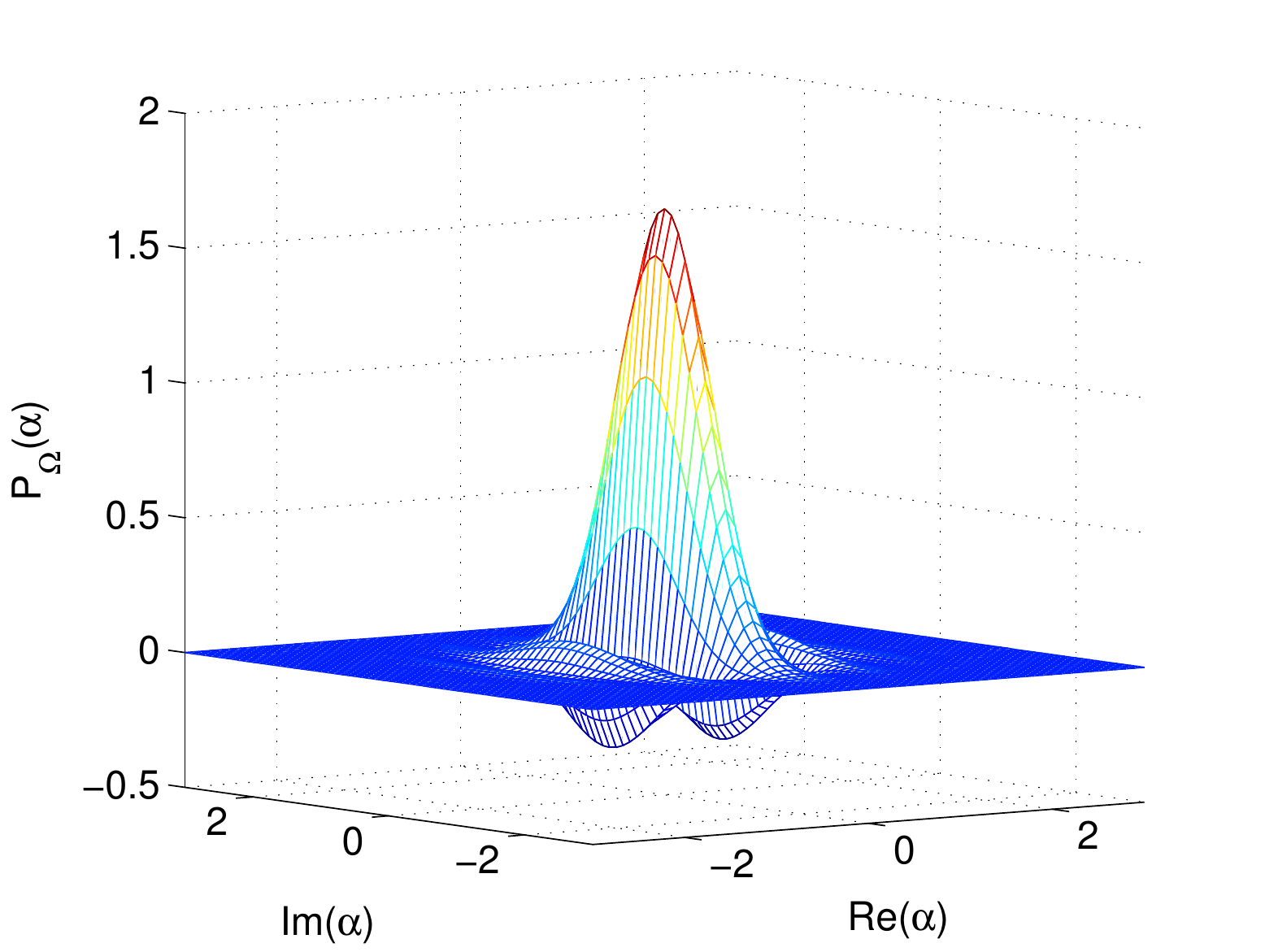}
	\caption{Nonclassicality quasiprobability of a squeezed state with $V_x = 0.5$ and $V_p = 2.0$.}
	\label{fig:POmega}
\end{figure}
\begin{figure}
	\includegraphics[width=0.8\columnwidth]{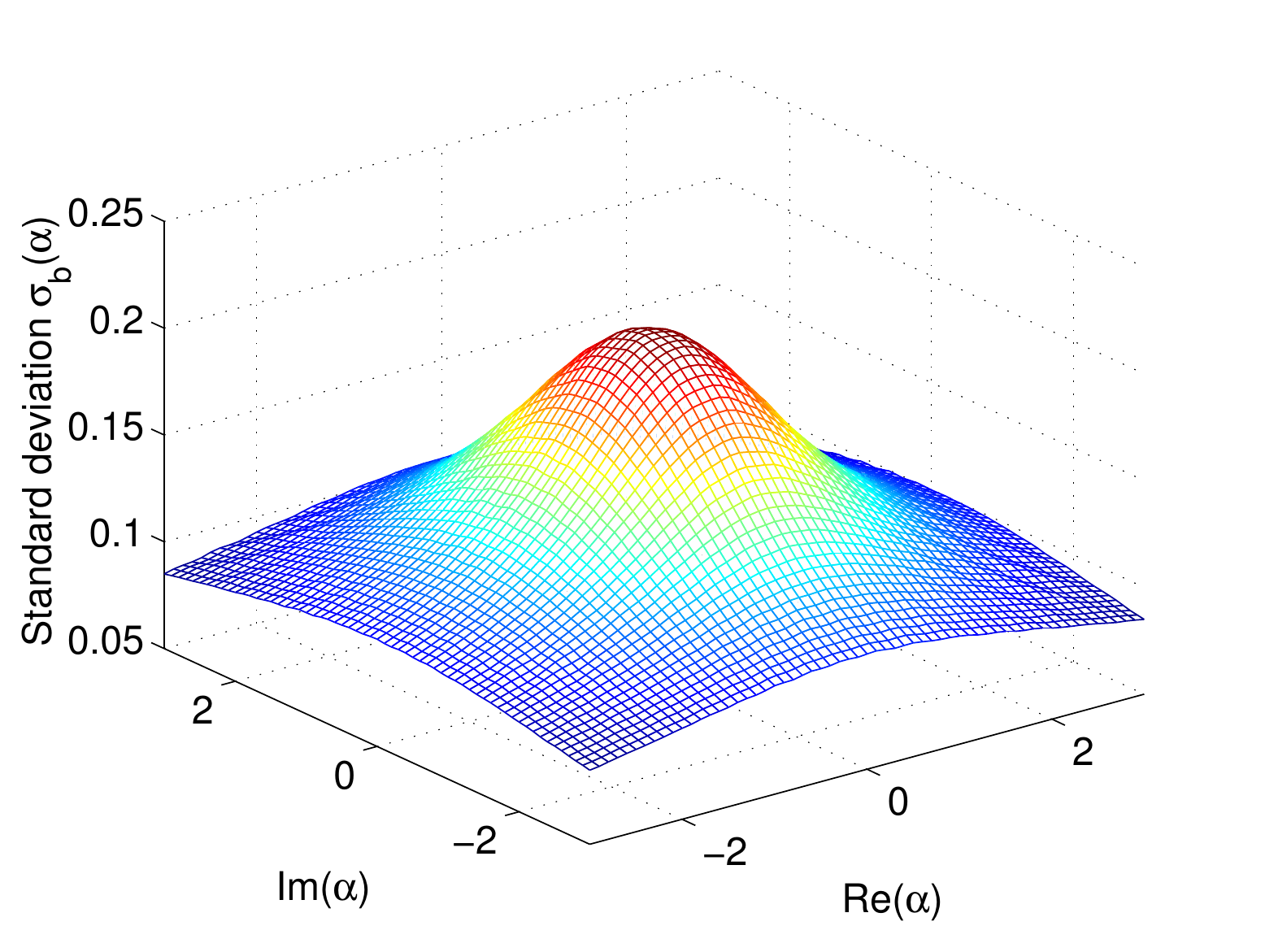}
	\caption{Standard deviation from balanced homodyne measurements of the nonclassicality quasiprobability in Fig.~\ref{fig:POmega}.}
	\label{fig:Stddev:Homo}
\end{figure}
\begin{figure}
	\includegraphics[width=0.8\columnwidth]{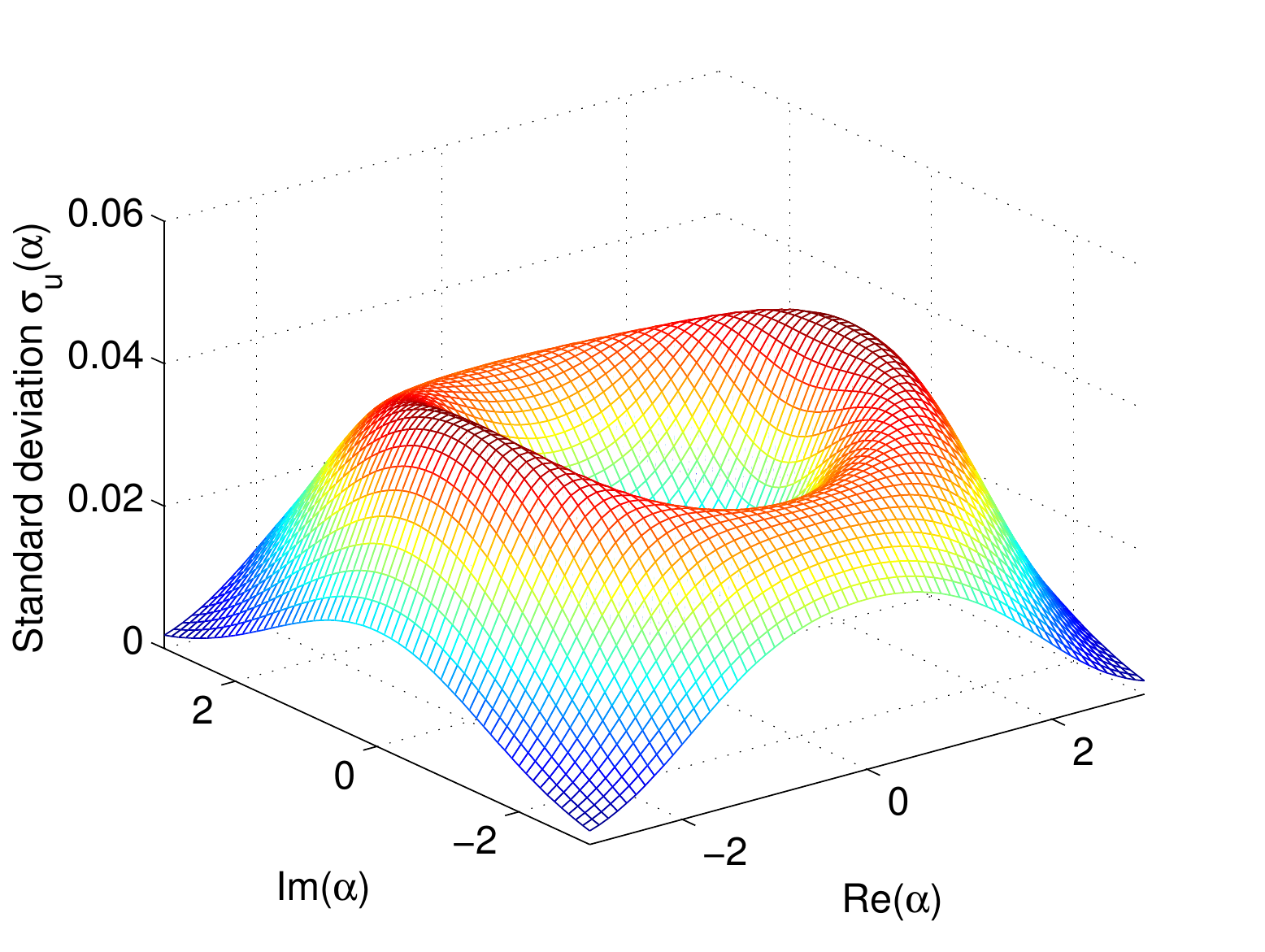}
	\caption{Standard deviation from unbalanced homodyne measurements of the nonclassicality quasiprobability in Fig.~\ref{fig:POmega}.}
	\label{fig:Stddev:Unbalanced}
\end{figure}
\begin{figure}
	\includegraphics[width=0.8\columnwidth]{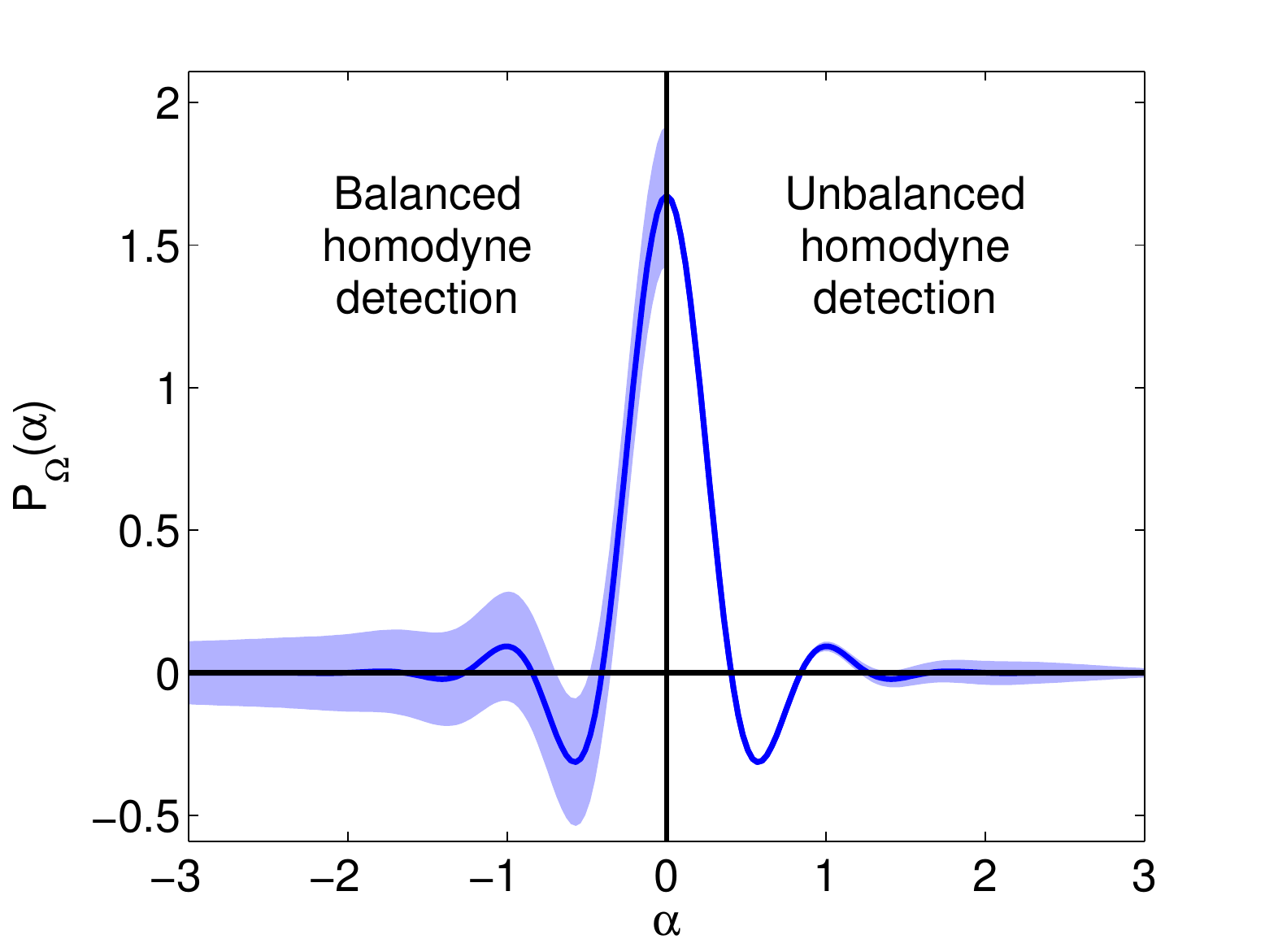}
	\caption{Comparison of the results for balanced and unbalanced homodyne tomography. The blue shaded area corresponds to one standard deviation.}
	\label{fig:Comparison}
\end{figure}

Figure~\ref{fig:POmega} shows the nonclassicality quasiprobability. We observe negativities for some real $\alpha$, being signatures of the nonclassicality of the squeezed state, the minimum is achieved at $\alpha = 0.6$ with $P(0.6) = -0.31$. Figures~\ref{fig:Stddev:Homo} and \ref{fig:Stddev:Unbalanced} show the standard deviations $\sigma_b(\alpha)$ and $\sigma_u(\alpha)$, which are obtained from balanced or unbalanced homodyne measurements respectively. They are calculated from Eq.~\eqref{eq:pattern:variance} and~\eqref{eq:var:F}, each divided by the number of measurements $N$. Obviously, they show a completely different behavior. The uncertainty from balanced homodyne detection is more than a factor of $3$ larger than the one from the unbalanced technique, the exact difference depends on the point in phase space and on the examined state. In particular, at $\alpha =0.6$, the homodyne measurement provides $\sigma_b(0.6) = 0.191$, leading to an insufficient significance of the negativity of $1.6$ standard deviations. Contrarily, we have $\sigma_u(0.6) = 0.010$ in the unbalanced case, leading to a significance of about $31$ standard deviations. Therefore, in case of equal quantum efficiency, the unbalanced scheme proves to be much better than the balanced one. Fig.~\ref{fig:Comparison} illustrates this conclusion clearly.

\section{Conclusion and Outlook}

We compared different approaches for estimating the expectation value of some physical quantity with respect to its statistical uncertainty. First, we showed that whenever one can estimate a quantity of interest as a function of a single operator, which can be directly measured, then the estimate is on the quantum mechanical level of uncertainty, i.e.~the empirical variance equals to the quantum-mechanical one. In practice, this works for quadratures and photon number measurements, for instance.

However, for many operators a direct measurement is not known, and techniques for quantum tomography have to be applied. We considered sampling methods, which are applied to phase-dependent quadrature data from balanced homodyne measurements. We show that the cascaded balanced homodyne measurement has no statistical advantage over the standard technique, although the first method only requires a phase-randomized local oscillator. Moreover, both methods do not operate on the quantum mechanical level of uncertainty. On the contrary, the unbalanced measurement technique can achieve the quantum mechanical level of uncertainty. Therefore, the latter shall be generally the best of the considered methods.

We also identified the main reason for the difference between the variance from balanced homodyne measurement and the quantum mechanical variance: It is due to the fact that quadratures at different phases are always measured stochastically independently. Therefore, there seems to be a lack of ``phase correlations'' in the experimental data. The severe question is how this problem affects other quantum state reconstruction methods, like maximum likelihood techniques~\cite{MaxLike1,MaxLike2}. In particular, it is unclear if the latter method is able to perform the estimation on the quantum mechanical level.

Our results have direct implications to the different approaches for reconstructing quasiprobabilities or density matrices of states. We showed the advantage of the unbalanced measurement with the example of a nonclassicality quasiprobability of a squeezed state. Therefore, if photon-counting devices with quantum efficiencies comparable to balanced homodyne detectors will be available in the future, our findings suggest to prefer the unbalanced scheme for quantum state estimation.

\section{Acknowledgments}
The author gratefully thanks W. Vogel for fruitful discussions and his assistance.
This work was supported by the Deutsche Forschungsgemeinschaft through SFB 652.

\end{document}